\documentclass[preprint,12pt]{elsarticle}
\usepackage{hyperref}
\usepackage{marginnote}
\usepackage{amsmath,amssymb,amsfonts}
\usepackage{algorithmic}
\usepackage{graphicx}
\usepackage{multirow}
\usepackage{textcomp}
\usepackage{booktabs}
\usepackage{float}
\usepackage{graphicx}
\usepackage{lipsum,booktabs}
\usepackage{adjustbox}
\usepackage{amssymb}
\usepackage{geometry}
\usepackage{xcolor}
\definecolor{snp}{HTML}{FDF38E}%{#FDF38E}
\definecolor{sgp}{HTML}{00B140}%{#00B140}
\definecolor{con}{HTML}{0087DC}%{#0087DC}
\definecolor{lab}{HTML}{E4003B}%{#E4003B}
\definecolor{lib}{HTML}{FAA61A}%{#FAA61A}
\definecolor{pc}{HTML}{005B54}% {#005B54}
\definecolor{sf}{HTML}{326760}% {#326760}
\definecolor{dup}{HTML}{D46A4C}% {#D46A4C}
\definecolor{sdlp}{HTML}{2AA82C}% {#2AA82C}
\definecolor{apni}{HTML}{F6CB2F}% {#F6CB2F}
\definecolor{tuv}{HTML}{0C3A6A}% {#0C3A6A}
\definecolor{pbp}{HTML}{E91D50}% {#E91D50}

\usepackage{subcaption,booktabs,multirow,comment,graphicx,lipsum} %nireak

\journal{arxiv}
\begin{document}

\begin{frontmatter}



\title{HTIM: Hybrid Text-Interaction Modeling for Broadening Political Leaning Inference in Social Media}


\author[inst1]{Joseba Fernandez de Landa}
\author[inst2]{Arkaitz Zubiaga}
\author[inst1]{Rodrigo Agerri}

\affiliation[inst1]{organization={HiTZ Center - Ixa, University of the Basque Country UPV/EHU}}
\affiliation[inst2]{organization={Queen Mary University of London}}

\begin{abstract}
Political leaning can be defined as the inclination of an individual towards certain political orientations that align with their personal beliefs. Political leaning inference has traditionally been framed as a binary classification problem, namely, to distinguish between left vs. right or conservative vs liberal. Furthermore, although some recent work considers political leaning inference in a multi-party multi-region framework, their study is limited to the application of social interaction data. In order to address these shortcomings, in this study we propose Hybrid Text-Interaction Modeling (HTIM), a framework that enables hybrid modeling fusioning text and interactions from Social Media to accurately identify the political leaning of users in a multi-party multi-region framework. Access to textual and interaction-based data not only allows us to compare these data sources but also avoids reliance on specific data types. We show that, while state-of-the-art text-based representations on their own are not able to improve over interaction-based representations, a combination of text-based and interaction-based modeling using HTIM considerably improves the performance across the three regions, an improvement that is more prominent when we focus on the most challenging cases involving users who are less engaged in politics.
\end{abstract}



\end{frontmatter}


\section{Introduction}\label{sec:introduction}

Ideology is broadly understood as a set of people's beliefs formed by ways of thinking and acting in society \citep{sartori1969politics}. Those beliefs can generally be represented by political parties, acting like social hubs of coordinated thoughts and actions. Thus, each region may have their own political parties, adjusted to a certain sociopolitical context in order to attract the votes and support of specific parts of the population. Understanding political leaning as the proximity to a political party, will allow to represent better ideological nuances than by reducing them to binary frameworks such as left vs right or conservative vs liberal. Therefore, our objective is to represent individual actors leveraging social media interactions and published text, allowing political leaning inference even in scenarios where no interactions are available, such as news media or political speeches. By doing so, we aim to incorporate wider and more accurate representations into the ongoing exploration of public opinions for diverse social science studies, including hate speech, disinformation or propaganda detection \citep{akhtar2019new,hristakieva2022spread}.

The task of inferring the political leaning of social media users has widely been tackled through the use of user interactions as features \citep{conover2011political,Barber2015UnderstandingTP,Barber2015BirdsOT,barbera2015tweeting,darwish2020unsupervised,garimella2017long,Wong2013QuantifyingPL}, primarily based on the assumption that the behaviour exhibited by users through interactions with one another reflect homophilic connections and patterns of polarization between users \citep{conover2011political,barbera2015tweeting,cinelli2021echo,10.1145/3178876.3186139}. Others have approached the task by leveraging textual data instead of interactions through the use of Natural Language Processing techniques \citep{PreotiucPietro2017BeyondBL,Kulshrestha2017QuantifyingSB,Pl2014PoliticalTI,Yan2019TheCC,PLN6446,Fagni2022FinegrainedPO,Rashed2021EmbeddingsBasedCF}. The few previous approaches that have combined both texts and interactions \citep{conover2011predicting,Pennacchiotti2011DemocratsRA,Lahoti2017JointNM}, have addressed the task as a binary classification problem, in which only two political dimensions (e.g. right vs left) are considered at a time \citep{conover2011political, conover2011predicting, Barber2015UnderstandingTP,Barber2015BirdsOT,garimella2017long, Pl2014PoliticalTI, PLN6446,darwish2020unsupervised,barbera2015tweeting,PreotiucPietro2017BeyondBL, Wong2013QuantifyingPL,Pennacchiotti2011DemocratsRA, Hua2020TowardsMA, Kulshrestha2017QuantifyingSB, Yan2019TheCC, Lahoti2017JointNM}. The few studies that have gone beyond binary classification have been limited to a single region \citep{Rashed2021EmbeddingsBasedCF,Fagni2022FinegrainedPO}. This binary and region-specific standpoint represents the political context as a static and uniform reality.

In previous research \citep{de2023generalizing}, we showed that the use of features derived from interactions between users can lead to high performance in inferring the political leaning of social media users that are actively engaged in politics. However, a pure interaction-based approach proved to have clear limitations in making predictions for users who are less engaged in politics, hence hindering the broader applicability of the approach. Taking this research as a starting point, we propose a hybrid approach that aims to make the most of both textual and interaction-based features characteristic of social media. To achieve this, we introduce Hybrid Text-Interaction Modeling (HTIM), which integrates textual and interaction-based information into a hybrid model for improved political leaning inference across broader groups of social media users of varying degrees of political engagement. In doing so, our work presents the first attempt at addressing political leaning inference across a range of multiclass political realities fusioning text and interaction data.

Our proposed HTIM approach is flexible in that it can incorporate different encoding techniques. We test HTIM with different methods including text-based approaches like TF-IDF, Word2Vec \citep{mikolov2013efficient}, and Transformers \citep{vaswani2017attention}, as well as interaction-based methods such as DeepWalk \citep{deepwalk}, Node2vec \citep{Grover2016node2vecSF} and Relational Embeddings \citep{fernandez2022relational}. We evaluate the resulting HTIM-based models in three datasets pertaining to three different regions of the UK, each with different political parties. We study the performance of our models on users with different levels of engagement in politics, with a particular focus on users with lower levels of engagement in politics, whose posting of political content and interactions with content and users relevant to politics are predicted to be less frequent, which poses an additional challenge. Broadening political leaning inference with users of lower levels of engagement is important when recent studies, such as a survey conducted by the UK government \citep{uberoi2022political}, show decreasing levels of engagement in politics and in democracy from certain demographic groups.

Our experiments demonstrate that while interaction-based representations are superior to those based on textual content only, fusioning both types of information using HTIM lead to improved results, particularly for users with lower levels of political engagement. We therefore demonstrate the potential to broaden political leaning inference to a wider group of users not necessarily limited only to users who are highly active in politics.

Our work makes the following novel contributions:
\begin{itemize}
\item We propose Hybrid Text-Interaction Modeling (HTIM), as a means for hybrid modeling of social media users leveraging textual and interaction-based features, showing that the combination of both data types is required for optimal performance, especially for users with lower political engagement. 
\item We enquire into the proposed HTIM framework based on multiple political leanings anchored on representative political parties, ensuring adaptability to diverse regions and different levels of political engagement.
\item We conduct political learning inference without relying on one specific data type. However, our results demonstrate that interaction-based Relational Embeddings outperform textual approaches, including those applying Transformer-based language models.
\item All the data resources such as labeled users or their texts and interactions will be made publicly available upon publication for further study.
\end{itemize}

\section{Related Work} 
In this section, we will discuss prior research on political leaning inference across different countries, paying special attention to the diverse data sources employed for this purpose. The analysis will commence with sources based on interaction-based data concerning retweets and user follows and subsequently advance to text-based data. Ultimately, we will consider approaches that integrate both types of data.

\subsection{Interaction-based Political Leaning Inference}

Well-known methodologies leverage interaction-based data obtained from social media for inferring political leaning. These methodologies are centered on user actions such as \textit{following} and \textit{retweeting}. For example, in one of the first research works about political leaning inference using Twitter political alignment was studied as a left-right spectrum in the US context, showing retweets as the most polarized interactions \citep{conover2011political}. \textit{Follow} interactions \citep{Barber2015UnderstandingTP,Barber2015BirdsOT} and even combinations of \textit{follow} and \textit{retweet} interactions \citep{garimella2017long} have also been leveraged to infer users left-right leaning. Similarly, retweets have also been used to infer the conservative and liberal leanings of Twitter users, quantifying such leanings \citep{Wong2013QuantifyingPL}, and even deploying them to analyze polarization and echo chambers \citep{barbera2015tweeting}. Additionally, other works have investigated \textit{retweet} behaviour in the context of pro- or anti-alignments among different topics \citep{darwish2020unsupervised}.

\subsection{Text-based Political Leaning Inference}

Other approaches to political leaning inference revolve around analyzing text-based data derived from social media. Several researchers have focused on linguistic textual features as a means to measure political leaning along the liberal-conservative spectrum \citep{PreotiucPietro2017BeyondBL}. Sentiment analysis has also been employed to study left-right political leaning, as demonstrated in Plà and Hurtado \cite{Pl2014PoliticalTI}. Moreover, text-based methods based on topic vectors \citep{Kulshrestha2017QuantifyingSB} or autoencoders trained on n-grams \citep{Yan2019TheCC} have been applied to explore users' alignment with respect to the Democratic and Republican parties. Rashed \textit{et al.} \cite{Rashed2021EmbeddingsBasedCF} used text-based embedding features to study political polarization in Turkey. 

Alternatively, the PoliticEs 2022 shared task \citep{PLN6446} aimed to extract political leaning from texts through an author profiling approach. In order to infer left or right alignment at user level, tweets published by the same author are grouped, either by concatenating them at the input stage or by merging the associated labels at the output stage. Best results in the PoliticEs 2022 task are achieved by methods based on Transformers \citep{vaswani2017attention}, mainly fine-tuning pre-trained models. A similar idea based on author profiling is proposed by Fagni and Cresci \cite{Fagni2022FinegrainedPO} for inferring political leaning on Italy focusing on political parties, achieving best results with word2vec \citep{mikolov2013efficient}.

\subsection{Hybrid Approaches to Political Leaning Inference}

Several studies have employed a combined approach, by mixing textual content with interactions-based data. A pioneering research combined and compared text and interaction-based data in order to capture the left-right leaning of Twitter users, showing that retweets where the most representative interaction type to capture political alignment \citep{conover2011predicting}. Another innovative study employed text and interactions, such as friends or followers, to infer the alignment of users with respect to the Democrat and Republican parties \citep{Pennacchiotti2011DemocratsRA}. Hua \textit{et al.} \cite{Hua2020TowardsMA} also infer Democrat or Republican party preference of users based on hashtags, \textit{retweets} and \textit{followers}. Continuing within the US framework, another study aimed to estimate user ideology as liberal or conservative, achieving best results while combining textual content with a social graph derived from retweets and followers \citep{Lahoti2017JointNM}. Finally, interaction and textual features were used to identify pro or anti independence Twitter users in Catalonia, Basque Country and Scotland \citep{Zubiaga2019PoliticalHI}.

\subsection{Related Tasks}

Similar to political leaning inference, but focused on a specific topic, stance detection tasks \citep{mohammad-etal-2016-semeval, hardalov-etal-2021-cross} commonly target the political domain. Recently released datasets \citep{cignarella2020sardistance,vaxxstance2021} enabled researchers to apply both textual content and social interactions \citep{TextWiller,QMUL-SDS,WordUp}, emphasizing the significance of interactions in determining the stance of users via manually engineered approaches. In previous research \cite{fernandez2022relational}, we introduced Relational Embeddings, achieving better results than DeepWalk \citep{deepwalk} and node2vec \citep{Grover2016node2vecSF} while showing that interaction and textual inputs provide best results across different datasets without any manual engineering. The present work leverages these and other methods within the HTIM framework for political leaning inference, thereby broadening their applicability to a wider spectrum of users of varying degrees of engagement in politics.

\section{Datasets}
\label{sec:datacol}

For building our own datasets, we follow a generalizable methodology applicable to different political realities \citep{de2023generalizing}. First, we choose the political context we want to analyze by selecting regions and their most relevant political parties. Then, we start with the data collection from Twitter by manually labeling users and collecting their associated textual and interaction-based data to build the proposed user representations. 

\subsection{Political Context}\label{sec:regsel}
Given our interest in exploring multi-party and diverse scenarios, we chose the United Kingdom (UK) and examined three regions with devolved governments: Scotland, Wales, and Northern Ireland. These regions possess a diverse political landscape, characterized by strong nationalist sentiments and a wide range of political options. Our analysis focuses on the major political parties represented in each region's devolved parliament. 

\textbf{Scotland (SCT):} The Scottish political landscape is dominated by five major parties, namely, the Scottish National Party (SNP), the Scottish Conservative \& Unionist Party (SCU), the Scottish Labour Party (SL), the Scottish Green Party (SGP), and the Scottish Liberal Democrats (SLD). The SNP is a center-left party that advocates for Scottish independence and prioritizes Scotland's membership in the European Union. The SCU, on the other hand, is a center-right party that supports the union with the UK and opposes Scottish independence. The SL is a center-left party that also supports the union with the UK and opposes Scottish independence. The SGP, a left-wing party, advocates for both Scottish independence and membership in the European Union. The SLD is a center party that supports both Scotland's membership in the European Union and the union with the UK.

\textbf{Wales (WAL):} In Welsh politics there are four major parties that dominate the landscape: Welsh Labour (WL), Welsh Conservatives (WC), Plaid Cymru (PC), and Welsh Liberal Democrats (WLD). WL is a center-left party that supports the UK and opposes Welsh independence. The WC is a center-right party also supporting the UK and opposing Welsh independence. PC is a left-leaning party that advocates for Welsh independence and greater autonomy from the UK. The WLD is a centrist party that supports the European Union and the union with UK.

\textbf{Northern Ireland (NIR):} In Northern Ireland politics, there are five main political parties: Sinn Féin (SF), Democratic Unionist Party (DUP), Alliance Party of Northern Ireland (APNI), Ulster Unionist Party (UUP), and Social Democratic and Labour Party (SDLP). SF is a left-wing party that advocates for Irish unity. The DUP is a right-wing party that supports the UK and opposes Irish unity. APNI is a centrist party that supports the UK and the European Union, and also advocates for greater cross-community cooperation in Northern Ireland. UUP is a center-right party that supports the UK and opposes Irish unity. The SDLP is a center-left party that supports Irish reunification and greater cooperation between Northern Ireland and the Republic of Ireland.

\subsection{Types of Users based on Levels of Political Engagement}

Political participation encompasses different levels of involvement \citep{almond2015civic}, ranging from passive observers to highly engaged activists, each harboring distinct perspectives and motivations. Therefore, in line with previous research in political science studying engagement levels \citep{ponce2013party}, in our work we define members, supporters, and sympathizers, labeled according to their alignment with a specific political party:

\begin{itemize}
 \item \textbf{Members} comprise users directly affiliated with political parties, including elected representatives and affiliated organizations, being users which are highly active in politics.
 \item \textbf{Supporters} are users closely aligned with political parties but not as actively engaged as members.
 \item \textbf{Sympathizers} are users with a loose connection to politics, making it more challenging to associate them directly with political parties due to their lower level of engagement which translates in very low activity related to political activities.
\end{itemize}

By incorporating different levels of political involvement we facilitate a more granular analysis of political landscapes, enhancing our comprehension of the multifaceted nature of political leaning. Additionally, we simulate realistic political scenarios to assess the methods in more feasible real-life situations. Member, supporter and sympathizer sets are designed to represent different levels of difficulty, with more involved user predictions expected to be easier than less involved user predictions. This step was taken to provide a more realistic evaluation of the proposed methods, facing situations that could happen in real life.

\subsection{Data Collection}\label{sec:datacolect}
Once the regions and the political parties are selected, we begin with the data collection process. The first step is to manually generate a seed dataset by labeling Twitter users from each of the identified parties. In a second step, we automatically extract users which are related to the already labeled users, categorizing them as interacting users. Finally, we extracted the timelines of all collected users to obtain considerable amounts of both textual and interaction-based data.

(1) \textbf{Labeling of Member users:} In line with previous studies \citep{Makazhanov2013PredictingPP, Barber2015BirdsOT, garimella2017long, Hua2020TowardsMA}, we employed a similar technique based on set of seed users to collect data for our study. We began by selecting a group of users who are affiliated with the political parties of interest, including elected members and associated organizations. These users were manually identified and labeled with the corresponding political party.  For each region, we curated a list of member users independently, which served as the basis for data collection. The number of user's labeled by class can be seen in Table \ref{tab:lab_augmented} (Member column).

(2) \textbf{Supporter and Sympathizer evaluation datasets:} We further gather two datasets for each area: one consisting of supporter users and a second one consisting of sympathizer users. To construct these evaluation sets, we first extracted users based on their political party affiliations and their followers as it has previously been done in similar approaches \citep{barbera2015tweeting, Rashed2021EmbeddingsBasedCF}. Thus, supporter users were defined as those who followed five or more member users from a specific political party, while sympathizer users followed two or fewer users from each party \citep{de2023generalizing}. We collected 100 users for each party from each region for both supporter and sympathizer sets, and these users were automatically labeled based on their party affiliation. Then we select users with available data to create the final datasets as seen in Table \ref{tab:lab_augmented} (Supporter and Sympathizer columns). 

\begin{table}[ht!]\small
\centering
\begin{tabular}{@{}lrrrr@{}}
\toprule
Region   & Party     &      Member  &    Supporter &   Sympathizer \\ \midrule
\multirow{6}{*}{SCT}     
&SNP \textcolor{snp}{$\bullet$} &181&91&74 \\             
&SCU \textcolor{con}{$\bullet$} &59&86&81  \\ 
&SL \textcolor{lab}{$\bullet$} &52&88&72  \\  
&SGP \textcolor{sgp}{$\bullet$} &42&82&77 \\      
&SLD \textcolor{lib}{$\bullet$} &24&90&84 \\ 
&total &358&437&388 \\ \midrule
\multirow{5}{*}{WAL}     
&WL \textcolor{lab}{$\bullet$}  &55&92&77\\               
&WC \textcolor{con}{$\bullet$}  &42&91&75\\      
&PC \textcolor{pc}{$\bullet$}  &42&91&72\\ 
&WLD \textcolor{lib}{$\bullet$}  &27&98&81\\  
&total &166&372&305\\  \midrule
\multirow{6}{*}{NIR}     
&SF \textcolor{sf}{$\bullet$}      &79&92&37 \\
&DUP \textcolor{dup}{$\bullet$}    &61&44&54 \\
&APNI \textcolor{apni}{$\bullet$}  &52&62&66 \\
&UUP \textcolor{con}{$\bullet$}    &57&52&48 \\
&SDLP \textcolor{sdlp}{$\bullet$}  &58&54&65 \\
&total &307&304&270\\\bottomrule
\end{tabular}
\caption{\footnotesize Manually labeled (Member) users and automatically labeled users (Supporter and Sympathizer) for realistic evaluation, by region and class.}
\label{tab:lab_augmented}
\end{table}

(3) \textbf{Timeline extraction:} In this phase we will get data to characterize the labeled twitter users based on text and interactions. Regarding textual data, we collected 120 tweets per Member user and 60 tweets for each Supporter and Sympathizer users. To achieve a balance between the number of users and available data, we excluded labeled users with insufficient data and discarded tweets with fewer than 10 tokens. In the case of interaction-based data, we first identified all the users who had interacted via retweets with each of the labeled users, referred to as \textit{interacting} users. After that all the available retweets were extracted from the timelines of both the labeled and interacting users.

Political party selection and manual labeling of Member datasets were carried out in September 2022. Supporter and Sympathizer users evaluation sets were built on October 2022. Twitter data extraction was undertaken during October 2022, collecting the timelines of all the identified users. In Table \ref{tab:data_final} it can be seen the shape of the final corpus for each region: (i) Manually labeled users (Members) and gathered tweets (120 per user); (ii) automatically labeled users for the realistic evaluation (Supporters and Sympathizers) and the collected tweets for each of them (60 per user); (iii) interaction based retweet data from labeled and interacting users.

\begin{table}[ht!]\small
\centering
\begin{tabular}{@{}llrrr@{}}
\toprule
            &&\textbf{SCT}&\textbf{WAL}& \textbf{NIR}   \\ \midrule
\multirow{3}{*}{\textbf{Members}} 
&users          & 358     & 166       & 307   \\ 
&tweets         & 42,960  & 19,920    & 36,840   \\
&tokens         & 1,400k  & 789k      & 653k   \\ \midrule
\multirow{3}{*}{\textbf{Supporters}} 
&users          & 437     & 372       & 304   \\ 
&tweets         & 26,220  & 22,320    & 18,240   \\
&tokens         & 654k    & 676k      & 497k   \\ \midrule
\multirow{3}{*}{\textbf{Sympathizers}} 
&users         & 388      & 305       & 270   \\ 
&tweets        & 23,280   & 18,300    & 16,200   \\
&tokens        & 1,194k   & 523k      & 436k   \\ \midrule \midrule
\multirow{2}{*}{\textit{Interactions}} 
&interacting users         & 87k      & 62k       & 21k    \\ 
&retweets      & 19M      & 21M       & 4M  \\ \bottomrule
\end{tabular}
\caption{\footnotesize Final dataset composition for each region.}
\label{tab:data_final}
\end{table}

\section{Methods}
We conduct experiments using various methods to extract user representations from text and interaction data. On the one hand, text-based features are employed to represent users using textual data. On the other hand, interaction-based features derived from retweets are utilized to compare them with the textual approaches. Finally we propose a combination of textual and interaction features in our HTIM approach. User representations are then used to perform political leaning inference via alignment to the political parties included in our dataset.

\subsection{Text-based Features}
We explore a range of text-based feature extraction techniques in order to represent users. The extraction process involves generating user vectors based on the textual content associated with each user, namely, their tweets. By utilizing tweets as input data, we aim to capture users' preferences and behaviors. With that purpose we will employ text-based user representations to predict users' political leaning, following similar methodologies to those employed in previous studies \citep{Fagni2022FinegrainedPO,fernandez2022relational,8875952}.

\paragraph{\textbf{Term Frequency Inverse Document Frequency}} 
The tfidf statistical measure assesses the relevance of a word to a document within a set of documents. By lowering the impact of words that occur too frequently in the selected text collection the most salient features are selected. Our use of this approach is motivated by the fact that a limited set of words significantly impact the final predictions \citep{shen-etal-2018-baseline} and the remarkable results obtained in other similar text classification approaches \citep{fernandez2022relational,PLN6446}. In this case, all the tweet collections from each user are considered as individual documents. The obtained tfidf vectors for each author or user are used to learn a classifier, proposing a user level classification.

\paragraph{\textbf{Word Embeddings}} 
We use Word2vec \citep{mikolov2013efficient} (w2v) to encode each of the tweets into text-based vector representations. To accomplish this we train our own models from scratch in order to fit all the words from our datasets into the vocabulary. A separate language model is trained for each region, aiming to capture the prevailing expressions within specific communities. These models are trained using default hyperparameters (C-BOW, negative sampling, 5 epochs, 5 words window size) but different dimensions are considered to select the optimal configuration. For our purposes, we extract tweet vectors one by one, representing each tweet as the average of its word vectors \citep{kenter-etal-2016-siamese, Fagni2022FinegrainedPO}.

\paragraph{\textbf{Transformers}} 
We use pre-trained Transformer models \citep{vaswani2017attention} as they have garnered considerable attention in recent years for text classification tasks, including user profiling through textual data \citep{PLN6446}. These models enable context and meaning representation by analyzing the relationships among tokens in a text sequence. Transformer-based contextualized embeddings values are modified depending on the surrounding words and their order, while static embedding methods such as word2vec represent words with fixed vector values. Four multilingual models have been selected for our experiments:

\begin{itemize}
 \item \textbf{mBERT} \citep{devlin-etal-2019-bert} model is the multilingual version of BERT \citep{devlin-etal-2019-bert} pre-trained with the largest 104 languages in Wikipedia. Rather than simply predicting the next word in the sequence, the BERT model takes into consideration all of the words in the sequence, thereby developing a more profound comprehension of the context. BERT pre-trains bidirectional representations from unlabeled text by considering both left and right context in all layers based on two pre-training objectives, namely, mask-language modeling and next sentence prediction.

 \item \textbf{DistilmBERT} \citep{sanh2019distilbert} as the multilingual version of DistilBERT is a smaller and faster Transformer model distilled from BERT, with 40\% fewer parameters and 60\% faster performance while maintaining over 95\% of BERT's performance on the GLUE benchmark.

 \item \textbf{XLM-RoBERTa} \citep{conneau2020unsupervised} is a multilingual version of RoBERTa-base \citep{liu2019roberta} pre-trained on a large multilingual corpus containing 100 languages. As a robustly optimized BERT approach, it is trained on a 10 times larger dataset than the used in BERT and using a dynamic masking technique, byte-pair encoding tokenization and without the next-sentence prediction objective. 

 \item \textbf{XLM-T} \citep{barbieri2021multilingual} is an extension of the XLM-RoBERTa base model further trained on 198 million multilingual tweets. Given its focus on Twitter-based data, it is particularly relevant to assess its performance in tasks that are specific to this social media platform.
\end{itemize}

To ensure consistency across all categories and prevent overfitting, we employ the Transformers models without fine-tuning, meaning that we use the default frozen weights as done in other approaches \citep{de2023hitz}. Text features are extracted separately for each tweet, treating each individual tweet as a sequence. We explored three distinct approaches for representing the text of tweets using transformers: (a) \textit{start-of-sequence} initial token embedding is used as the entire tweet representation \citep{devlin-etal-2019-bert}; (b) \textit{average} value of the output embeddings of all words in the tweet to represent each tweet as the average of its word vectors \citep{kenter-etal-2016-siamese, Fagni2022FinegrainedPO}; (c) \textit{max-pool} value of all the words in a tweet to extract the most salient features from every word-embedding, by taking the maximum values among all the word vectors \citep{zhelezniak2018dont}.

\begin{figure*}[ht!] \footnotesize
  \centering
    \includegraphics[width=0.8\linewidth]{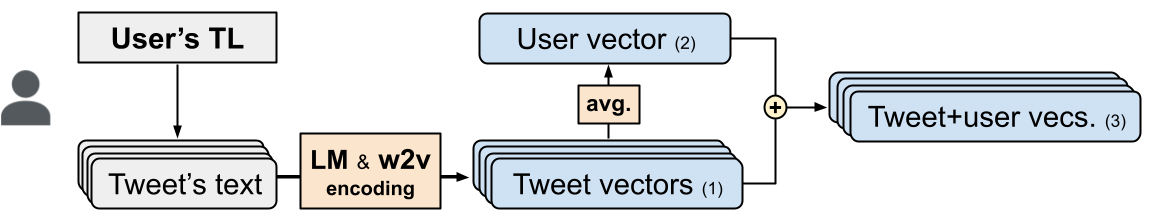}
  \caption{\footnotesize Pre-trained Transformer-based Language Models (LM) and Word2vec embeddings (w2v) usage to extract text-based user representations: (1) at Tweet level, (2) at user level and (3) the combination between tweet and user features.}
  \label{fig:tweet&user}
\end{figure*}

While the tfidf method is capable of directly representing all of a user's content at once through user-level features, Word2vec and Transformers are able to extract information only at the tweet level. As the tweet level information is not sufficient to represent users due to the small amount of text, we generate and add user-level textual features to each of the tweets representation as shown in Figure \ref{fig:tweet&user}. So, we obtain the textual representations of all the tweets for each user (1) which are then concatenated and averaged to obtain a user-based vector representation (2), emulating similar approaches \citep{8875952, kenter-etal-2016-siamese, le2014distributed, Rashed2021EmbeddingsBasedCF}. The final representation (3) consists of the concatenation of each tweet vector with the user-based vector representation. The combined user and tweet representations for each tweet are used to train a classifier. After predicting the labels at tweet level, a majority voting strategy is employed to infer the user label by considering the various tweet labels associated with the same author \citep{PLN6446,de2023hitz}.

\subsection{Interaction-based Features}
The intuition is that of building interaction based-user representations able to capture sociopolitical information, without any textual data. We focus on retweets as they have been previously proven effective to perform user classification \citep{conover2011predicting, magdy16, darwish2020unsupervised, Stefanov2020PredictingTT, fernandez2022relational}. User interactions are brought into a low dimensional vector space, modeling retweets into user level representations. Three distinct unsupervised techniques are used to represent users, namely DeepWalk \citep{deepwalk}, Node2vec \citep{Grover2016node2vecSF} and Relational Embeddings \citep{fernandez2022relational}. Other approaches such as GCN \citep{Kipf2017SemiSupervisedCW}, GAT \citep{Velickovic2018GraphAN} or TIMME \citep{timme20} are discarded as they are not suitable for unsupervised learning and have high memory requirements. 

\textbf{DeepWalk} (DW) algorithm simulates random walks among connected nodes in a network to learn feature representations.  It predicts the context or neighbors of an instance using the Skip-gram method \citep{mikolov2013efficient}. The context is generated by random walks among surrounding connected data points, with the length and number of walks determining the context. 

\textbf{Node2vec} (N2V) method, similar to DeepWalk, introduces two parameters, $p$ and $q$, to influence the network structure during random walks. The return parameter ($p$) determines the likelihood of revisiting nodes and the in-out parameter ($q$) controls the probability of exploring unexplored graph areas.

\textbf{Relational Embedings} (RE) method aims to predict which user retweeted another user among all the collected interaction pairs. Unlike generating random walks among nearest neighbors, this approach focuses on the real relationships between two users.

We use all identified users, including those labeled and interacting users, together with their retweets, to provide input for the techniques mentioned above for training the models. Following previous work \citep{darwish2020unsupervised,Stefanov2020PredictingTT,fernandez2022relational}, by applying any of the three methods above we obtain low dimensional dense interaction-based representations. The user-based features are generated without applying any data filtering to any of the models. In order to create meaningful features while keeping the dimensionality low, we set the feature dimensions to 20 for DeepWalk, node2vec, and Relational Embeddings, drawing from previous studies \citep{darwish2020unsupervised,Stefanov2020PredictingTT,fernandez2022relational}. For node2vec and DeepWalk, we use the default parameter values typically employed by these algorithms: walks\_per\_node = 10, walk\_length = 80, window or context\_size = 10, and one epoch \citep{deepwalk,Grover2016node2vecSF}. Specifically for node2vec, we set p=1 and q=0.5 to emphasize network community-related information \citep{Grover2016node2vecSF}.

\begin{table*}[ht!]\scriptsize
\centering
\begin{tabular}{@{}ll|llll|llll|llll|llll@{}}
\toprule
&& \multicolumn{4}{c}{SCT}  & \multicolumn{4}{c}{WAL}  & \multicolumn{4}{c}{NIR}  & \multicolumn{4}{c}{average} \\ 
&Dims. &50&100&200&300 &50&100&200&300 &50&100&200&300 &50&100&200&300 \\ \midrule
& tfidf    
& 22.4 & 39.3 & 48.4 & \underline{57.2}
& 45.3 & 54.8 & 64.2 & \underline{64.8}
& 30.3 & 40.4 & 55.5 & \underline{60.2} 
& 32.7 & 44.8 & 56.0 & \underline{60.7} \\
& w2v   
& 57.9 & 62.3 & 63.5 & \underline{64.0}   
& 61.8 & 61.3 & 61.3 & \underline{64.0}   
& 54.2 & 56.3 & 57.5 & \underline{57.9} 
& 58.0 & 60.0 & 60.8 & \underline{62.0} \\
\bottomrule
\end{tabular}
\caption{\footnotesize F1 macro score results 10 fold CV on SCT, WAL and NIR Members datasets. Algorithms used to generate the features: tfidf and w2v. Underlined values represent best result for each algorithm in each dataset.}
\label{tab:f1_results_base}
\end{table*}

\begin{table*}[ht!]\scriptsize
\centering
\begin{tabular}{@{}ll|llll|llll|llll@{}}
\toprule
&& \multicolumn{4}{c}{SCT}  & \multicolumn{4}{c}{WAL}  & \multicolumn{4}{c}{NIR}   \\ 
&Transformers&B&dB&R&Rt  &B&dB&R&Rt &B&dB&R&Rt \\ \midrule
\multirow{3}{*}{Features}     
& start-of-sequence   
& 35.9      & 33.7      & 07.2      & 37.2    
& 55.7      & 55.3      & 41.0      & 56.4    
& 45.6      & 42.0      & 10.4      & 39.0 \\ 
& average  
& 54.2      & 48.6      & 12.7      & 39.7     
& 68.4      & 64.0      & 52.0      & 56.1     
& 56.7      & 49.3      & 20.1      & 40.1 \\ 
& max-pool  
& \underline{67.6}      & \underline{73.4}      & \underline{47.6}      & \underline{60.7}     
& \underline{75.6}      & \underline{75.1}      & \underline{58.0}      & \underline{64.4}    
& \underline{68.4}      & \underline{72.1}      & \underline{50.7}      & \underline{57.4} \\ 
\bottomrule
\end{tabular}
\caption{\footnotesize F1 macro score results for 10 fold CV on SCT, WAL and NIR Members datasets. Algorithms used to generate the features: mBERT (B), DistilmBERT (dB), XLM-RoBERTa (R), XLM-T (Rt). Underlined values represent best results for each algorithm in each dataset.}
\label{tab:f1_results_trans}
\end{table*}

\begin{figure*}[ht!] \footnotesize
  \centering
    \includegraphics[width=0.8\linewidth]{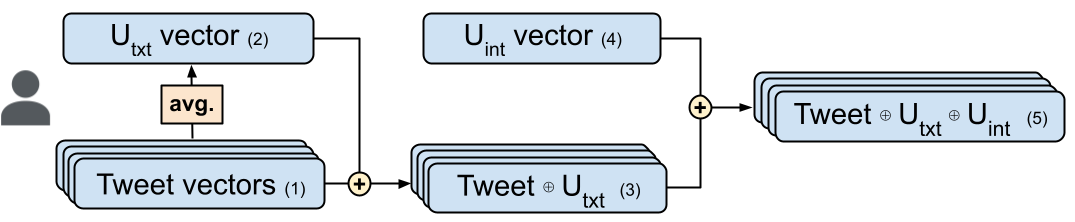}
  \caption{\footnotesize Hybrid Text-Interaction Modeling (HTIM) for each user tweet by tweet: (1) Tweet level text representation, (2) user level text representation, (3) the combination between tweet and user text features, (4) user level interaction representation and (5) final hybrid representation concatenating all vectors.}
  \label{fig:htim}
\end{figure*}

\subsection{HTIM: Hybrid Text-Interaction Modeling}

We propose a hybrid approach which integrates both the textual content expressed by a user and their corresponding social media interactions. When using text representation methods at tweet level (Figure \ref{fig:htim}), the textual representation of the tweets for each of the users (1) are averaged to obtain a vector characterizing each user (2) and concatenated with each of the tweets written by the author (3). Afterwards, (4) we concatenate them with the interaction-based representations to obtain a final (5) hybrid feature for each of the tweets. Those hybrid representations are used to train a classifier and to predict the labels at tweet level, and we subsequently use a majority voting strategy to infer the user label by considering the various tweet labels associated with the same author \citep{PLN6446,de2023hitz}.

Alternatively, when text representation methods (i.e. tfidf) facilitate the extraction of text features at the user-level, hybrid features are created at user level, concatenating user-level text and interaction features. User-level hybrid representations are used to train a classifier and to predict the labels at user level.

\section{Experimental Setup}
We leverage the obtained representations in order to conduct political leaning inference. We do so in two different sets of experiments. First, we focus on determining the optimal configuration to obtain good quality textual representations. Second, we will apply the best textual representations, together with interaction-based representations, in our HTIM hybrid approach, evaluating them on the Member, Supporter and Sympathizer datasets across the 3 regions.

\paragraph{\textbf{Textual Methods Selection}} 
For each region, we experiment with the users in the group of \emph{Members} using a 10 fold cross-validation (CV) setting, i.e., a 10\% of the users are left for evaluation while all the others are used for training. The primary objective of this experiment is to compare the performance of user representation methods. The representations obtained using the different methods are used to train a RBF-kernel Support Vector Machine (SVM) classification algorithm. We use the scikit-learn implementation \citep{scikit-learn} with default configuration.

With respect to tfidf and w2v, different dimension values were tested, as shown in Table \ref{tab:f1_results_base}. Ultimately, the best dimension value for both methods is set to 300. Particularly, for tfidf, as the dimension value increases, the results tend to improve. Regarding Transformer-based methods, the results in Table \ref{tab:f1_results_trans} demonstrate that \textit{max-pooling} proves to be the most effective strategy. The best textual user representation methods are used to be compared to and combined with interaction-based user representation methods.

\paragraph{\textbf{Experimental Settings}}
We will extract representations from tweets and interactions to train independent user classification models for each of the political regions: SCT, WAL and NIR. Each of the regions will have its own user representations to observe the performance of the methods on different scenarios.

In order to compare the quality of our proposed user representation methods considering the level of political engagement of the users, we evaluate separately on the Member, Supporter and Sympathizer datasets. On the one hand, we train and evaluate a SVM (RBF kernel) classifier \citep{scikit-learn} using 10 fold CV with the Members dataset. On the other hand, we use the Members dataset to train another SVM (RBF kernel) classifier and then evaluate the model on the Supporter and Sympathizer datasets. To ensure consistency across all categories and prevent overfitting, we have employed the default or automatic hyperparameters for all the aforementioned classifiers. In addition, majority and random label predictors are added as baselines for each of the datasets.

\begin{table*}[ht!] \tiny
\centering
\begin{tabular}{@{}ll|lll|lll|lll|lll|l@{}} 
\toprule
&& \multicolumn{3}{c}{SCT}  & \multicolumn{3}{c}{WAL}  & \multicolumn{3}{c}{NIR} & \multicolumn{3}{c}{average} & \\ 
&&Mem.&Sup.&Sym.            &Mem.&Sup.&Sym.           &Mem.&Sup.&Sym.           &Mem.&Sup.&Sym.   &ALL \\ \midrule
\multirow{2}{*}{Baselines}
& majority 
& 13.4  & 06.9  & 06.4      & 12.4  & 09.9  & 10.1     & 08.2  & 09.3  & 04.8   & 11.3 & 08.7 & 07.1 & 09.0  \\
& random 
& 17.8  & 21.6  & 18.7      & 21.5  & 27.9  & 28.1     & 21.2  & 21.9  & 16.1   & 20.2 & 23.8 & 21.0 & 21.6  \\  \midrule
\multirow{3}{*}{Interactions}
& RE   
& \underline{99.4}  & \underline{91.5}  & \underline{62.7}     & \underline{97.9} & \underline{95.8}  & \underline{59.6}     & \underline{97.6}  & \underline{72.9}  & \underline{33.0}    &\underline{98.3}&\underline{86.7}&\underline{51.8}&\underline{78.9}\\
& N2V   
& 80.8  & 62.0  & 17.4      & 67.5  & 48.5  & 11.1     & 60.8  & 21.6  & 07.0    & 69.7 & 44.0 & 11.8 & 41.9  \\
& DW  
& 80.9  & 61.6  & 22.9      & 77.5  & 57.0  & 14.7     & 72.2  & 23.6  & 06.6    & 76.9 & 47.4 & 14.7 & 46.3  \\ \midrule
\multirow{6}{*}{Text}
& tfidf 
& 57.2  & 52.4  & 29.5      & 64.8  & 59.5  & 26.1     & 60.2  & 45.8  & 25.9    & 60.7 & 52.6 & 27.2 & 46.8  \\ 
& w2v
& 64.0  & 40.5  & 27.5      & 64.0  & 51.2  & 31.5     & 57.9  & 37.9  & 26.7     & 62.0 & 43.2 & 28.6 & 44.6  \\
& B 
& 67.6  & 47.8  & 27.7      & \underline{75.6}  & 55.2  & \underline{33.9}     & 68.4  & 48.5  & \underline{34.4}    & 70.5 & 50.5 & 32.0 & 51.0  \\
& dB 
& \underline{73.4}  & \underline{59.0}  & \underline{36.9}      & 75.1  & \underline{64.1}  & 33.6     & \underline{72.1}  & \underline{49.8}  & 28.5       &\underline{73.5}&\underline{57.6}&\underline{33.0}&\underline{54.7}  \\
& R 
& 47.6  & 43.7  & 29.8      & 58.0  & 42.0  & 28.0     & 50.7  & 42.0  & 24.3    & 52.1 & 42.6 & 27.4 & 40.7  \\
& Rt 
& 60.7  & 48.3  & 34.5      & 64.4  & 52.2  & 32.7     & 57.4  & 44.0  & 27.8    & 60.8 & 48.2 & 31.7 & 46.9  \\ \midrule
\multirow{18}{*}{HTIM}
& RE + tfidf 
& \textbf{99.7}* &  \textbf{97.7}* &  \textbf{74.2}* & \textbf{99.2}* & \textbf{98.4}* & \textbf{67.0}* & 97.3  & \textbf{82.8}* & \textbf{48.1}*    &98.7*&\textbf{93.0}*&\textbf{63.1}*&\textbf{84.9}*\\ 
& RE + w2v
& 99.4  & 95.6* & 66.0*     & 98.5* & 96.0* & 64.4*            & 98.2* & 72.7  & 37.2*              &98.7*&88.1*&55.9*&80.9*\\
& RE + B  
& 98.5  & 94.0* & 63.1*     & \textbf{99.2}* & 94.9 & 60.8*     & 97.7* & 78.5* & 45.8*             &98.5*&89.1*&56.6*&81.4*\\
& RE + dB 
& 99.4  & 95.4* & 64.3*     & \textbf{99.2}* & 94.7 & 61.6*     & \textbf{98.4}* & 80.2* & 42.9*    &\textbf{99.0}*&90.1*&56.3*&81.8*\\
& RE + R  
& 99.4  & 94.6* & 66.0*     & 98.6* & 96.5* & 64.7*             & 97.4  & 78.5* & 41.1*             &98.5*&89.9*&57.3*&81.9*\\
& RE + Rt 
& 99.4  & 94.4* & 66.3*     & \textbf{99.2}* & 96.0* & 62.3*    & 98.1* & 78.4* & 44.7*             &98.9*&89.6*&57.8*&82.1*\\
\cmidrule{2-15}
& N2V + tfidf 
& 74.5 & 44.5 & 11.3      & 42.2 & 26.7 & 11.8      & 32.5 & 10.3 & 06.7   & 42.8	& 34.6 & 11.1 & 29.5 \\ 
& N2V + w2v
& 89.8* & 56.0  & 28.6*     & 75.1* & 60.2* & 32.7*    & 73.1* & 45.2* & 25.4                 & 79.3*& 53.8*& 28.9*& 54.0* \\
& N2V + B 
& 77.6  & 52.1  & 27.7      & 74.5  & 55.9  & 35.6*    & 68.0  & 48.3  & 34.6*                & 73.4*& 52.1*& 32.6*& 52.7* \\
& N2V + dB 
& 83.2* & 60.5  & 35.0      & 74.8  & 65.1* & 34.4*    & 71.6  & 50.1* & 29.2*                & 76.5*& 58.6*& 32.9 & 56.0* \\
& N2V + R 
& 75.4  & 47.1  & 28.2      & 58.4  & 45.6  & 26.7     & 52.5  & 37.3  & 22.6                 & 62.1 & 43.3*& 25.8 & 43.8* \\
& N2V + Rt 
& 79.4  & 51.0  & 32.8      & 65.5  & 57.6* & 30.6     & 57.9  & 42.8  & 28.1                 & 67.6 & 50.5*& 30.5 & 49.5* \\
\cmidrule{2-15}
& DW + tfidf
& 71.9 & 38.1 & 13.1      & 57.1 & 28.8 & 10.5      & 45.7 & 13.2 & 06.7   & 46.5	& 37.5 & 12.3 & 32.1 \\ 
& DW + w2v
& 87.5* & 58.0  & 30.1*     & 79.4* & 64.2* & 31.0     & 78.5* & 46.2* & 29.2*                & 81.8 & 56.1*& 30.1*& 56.0* \\
& DW + B  
& 76.5  & 53.3  & 27.5      & 75.5  & 56.4  & 35.6*    & 69.2  & 49.2* & 34.3                 & 73.7 & 53.0*& 32.5*& 53.1* \\
& DW + dB  
& 83.8* & 62.0* & 35.6      & 75.9  & 65.0* & 33.9*    & 72.1  & 50.2* & 29.4*                & 77.3*& 59.1*& 33.0*& 56.4* \\
& DW + R 
& 74.6  & 43.4  & 25.1      & 63.9  & 50.5  & 26.8     & 62.0  & 37.9  & 22.0                 & 66.8 & 43.9*& 24.6 & 45.1* \\
& DW + Rt 
& 78.6  & 48.9  & 33.5      & 67.8  & 57.3  & 29.8     & 61.8  & 42.4  & 27.7                 & 69.4 & 49.5*& 30.3 & 49.8* \\
\bottomrule
\end{tabular}
\caption{\footnotesize Macro-F1 scores on SCT, WAL and NIR from Member (10 fold CV), Supporter and Sympathizer datasets. Algorithms used to generate the representations: Relational Embeddings (RE), Node2vec (N2V), DeepWalk (DW), tfidf, word2vec (w2v), mBERT (B), DistilmBERT (dB), XLM-RoBERTa (R), XLM-T (Rt). Values in \textbf{bold} represent best overall results for each dataset, while \underline{underlined} values represent best results on text-only and interactions-only framework. Values with * represent when the combination of text and interactions gets better results than each on its own.}
\label{tab:f1_results}
\end{table*}

\section{Analysis of Results}
In this section we analyze the results obtained for different regions (SCT, WAL and NIR) on Member, Supporter and Sympathizer datasets, testing the methods through regions and different levels of political attachment. On the one hand, we are interested in evaluating standalone text-based and interaction-based methods. On the other hand, we compare standalone methods with the use of HTIM for combining interactions and text.

\paragraph{\textbf{Text vs interaction representations}} 
In Table \ref{tab:f1_results} we can compare the results for text and interaction-based representations on their own, showing which data type is better to represent political leaning. Regarding text-based representations, Bert-based (B and dB) approaches generally yield superior results compared to Roberta-based (R and Rt), w2v and tfidf methods. Interestingly, despite being a smaller model, dB achieves superior results compared to B, being the best text based approach. Among Roberta-based approaches, Rt is significantly superior to the R method, given that the former is an extended version of R that has been retrained using Tweets. Bert-based (B and dB) approaches also outperform interaction-based N2V and DW on politically less engaged users. However, interaction-based approaches tend to be better with politically engaged Member users.

In conclusion, max-pooled DistilmBERT Transformer model is the best approach to tackle political leaning inference with textual data, outperforming other Transformer configurations as well as w2v and tfidf baselines. Furthermore, we have to remark that none of the textual approaches come close to surpassing the performance of interaction-based RE representations, which consistently yield superior results across all regions and political attachments. However, RE is not able to perform well with politically less engaged users, as the performance drops as the political involvement decreases.

\paragraph{\textbf{Representations using HTIM}}
Taking a wider look into the whole Table \ref{tab:f1_results}, the results indicate that the use of our proposed HTIM for the integration of text and interaction representations leads to a notable improvement in results as compared to their independent usage (ALL column on Table \ref{tab:f1_results}). The RE method consistently yields better results when compared to other approaches based on interactions or text on their own. However, incorporating any of the extracted text representations alongside the RE representations often leads to improved results. This is particularly beneficial for politically less engaged users since their interactions alone may not provide enough information to determine their orientation accurately. 

Thus, RE combined with Transformer-based representations (B, dB, R, Rt) generally perform better than when the RE are on their own. Especially, RE combined with Rt (RE+Rt) outperformed RE method for the whole 3 regions and all the political engagements. Hence, the combination of RE with Transformer-based representations (B, dB, R, Rt) typically yields superior performance compared to standalone RE. Furthermore, tfidf representations, which do not perform well when combined with \textit{DW} and \textit{N2V} representations, considerably enhance the results achieved by \textit{RE} when combined with them. Thus, results on the Sympathizer datasets improve more than 10 points in average across the three regions, while for Supporter they increased more than 5 points. Considering that the added \textit{tfidf} representations are based on the most significant terms per user, we can hypothesize that certain referential terms may serve as anchor terms for specific political parties.

\begin{figure}[ht!]
\centering
    \includegraphics[width=0.33\linewidth]{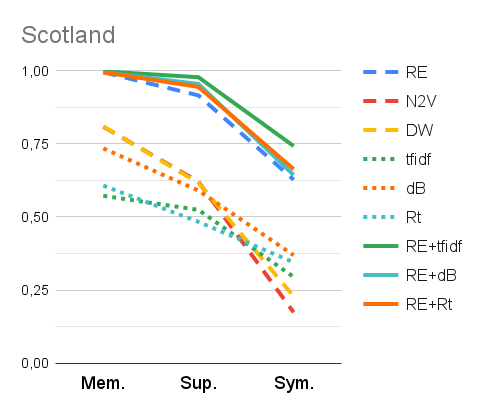}\hfil 
    \includegraphics[width=0.33\linewidth]{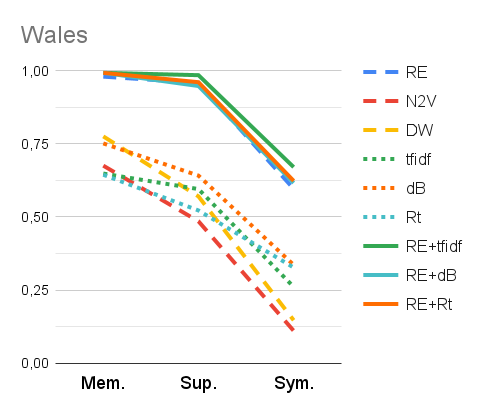}\hfil 
    \includegraphics[width=0.33\linewidth]{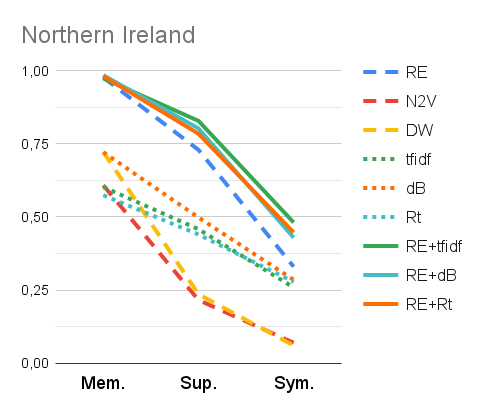}
    \caption{\footnotesize Performance variations for interaction-based approaches (RE, N2V and DW), best text-based approaches (tfidf, dB and Rt) and corresponding HTIM approaches (RE+tfidf, RE+dB and RE+Rt) among different levels of political engagement on SCT (left), WAL (center) and NIR (right) datasets.}
    \label{fig:results_enga} 
\end{figure}

\paragraph{\textbf{Results for different levels of engagement}}
We next look at the performance scores with a focus on the three levels of engagement, i.e. members, supporters and sympathizers. Results for these three groups as shown in Figure \ref{fig:results_enga} indicate that, as hypothesized, politically less engaged users are more difficult to predict for all the selected approaches. This observation is supported by the downward trend of all the lines, visually showing a steep performance decrease as engagement fades. This trend in turn reinforces the need for a data collection strategy like the one we defined here to collect not only actively engaged users, but also those with more modest levels of engagement which do need to be considered in these analyses. Furthermore, we also observe that political leaning inference of highly engaged users is best achieved by exploiting their content sharing actions or interactions (N2V and DW); this however changes when we shift our focus towards less engaged users, where the use of textual content becomes more crucial as the interactions alone do not suffice (i.e. tfidf, dB and Rt). Interestingly, the combination of both data types through HTIM further improves the results, especially when the hard-to-beat RE method underperforms on politically less engaged users. 

\begin{figure}[ht!]
\centering
    \includegraphics[width=0.33\linewidth]{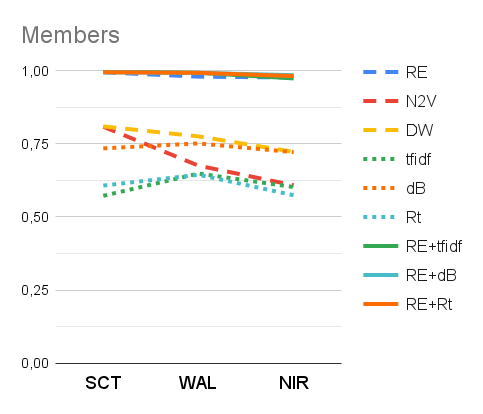}\hfil 
    \includegraphics[width=0.33\linewidth]{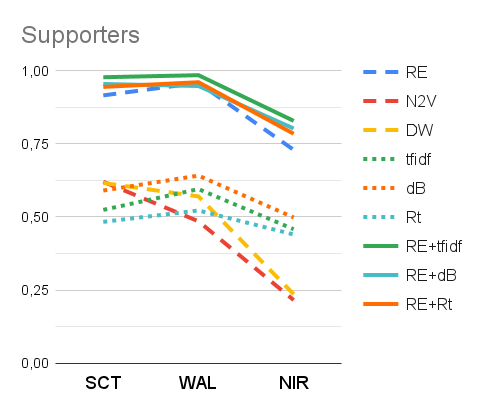}\hfil 
    \includegraphics[width=0.33\linewidth]{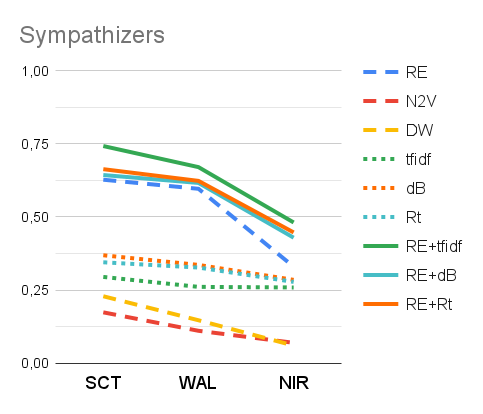}
    \caption{\footnotesize Performance variations for interaction based approaches (RE, N2V and DW), best text based approaches (tfidf, dB and Rt) and corresponding HTIM approaches (RE+tfidf, RE+dB and RE+Rt) among regions for members (left), supporters (center) and sympathizers (right) datasets.}
    \label{fig:results_region} 
\end{figure}

\paragraph{\textbf{Results across regions}}
We are also interested in looking at how results differ across the three regions under study, i.e. Scotland (SCT), Wales (WAL) and Northern Ireland (NIR). We show results for each region in Figure \ref{fig:results_region}, which shows that performance scores across regions also vary depending on the level of political engagement. Performance is consistenly high and comparable across all three regions when we look at highly engaged users (i.e. members). These trends vary more across regions when we look at less engaged users. As performance decreases for less engaged users across all three regions, we see that this drop is more prominent to some extent for WAL but particularly for NIR. Performance is more stable across regions when text-based methods (tfidf, dB and Rt) are used than when interaction-based methods (N2V and DW) are used, given that the availability of interaction data varies across regions. The weakest overall results occur within the NIR region, not least when interaction-based approaches are used on less engaged users. These weak results are however mitigated through the use of HTIM, especially when used in combination with RE+tfidf, which again proves to be a more robust strategy to be used, both to ensure consistency across regions as well as to better generalize on less engaged users.

\section{Discussion}
In this section we will perform a more comprehensive examination of the reported findings. We will do this by conducting an error analysis and creating visual representations of user data for the most effective method, namely, HTIM RE+tfidf.

\paragraph{\textbf{Error analysis}}
Interaction-based RE representations have a hard-to-beat performance, but when combined with text representations, they perform even better.  Thus, when combining any textual representations with RE representations, they demonstrate comparable or superior performance. Interestingly, transformer-based approaches are not the optimal representations to combine with RE. The most effective representations for combination are tfidf representations. HTIM representation of RE and tfidf (RE+tfidf) yielded significantly improved results, particularly for SCT and NIR Sympathizer users, surpassing the F1-score of standalone RE by more than 10 points. To visualize the improvement, confusion matrices are plotted for these datasets.

\begin{figure}[ht!]
\centering
    \includegraphics[width=0.38\linewidth]{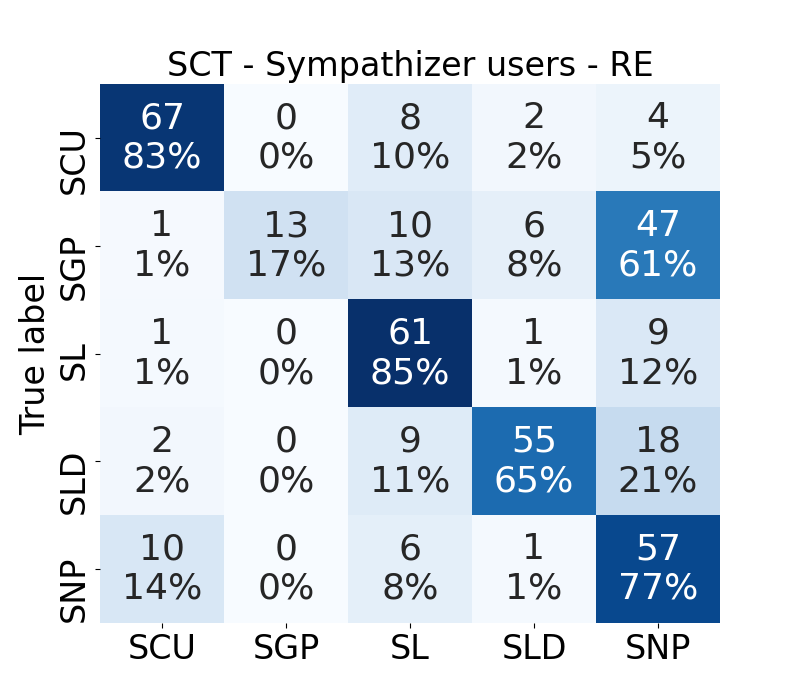}\hfil 
    \includegraphics[width=0.38\linewidth]{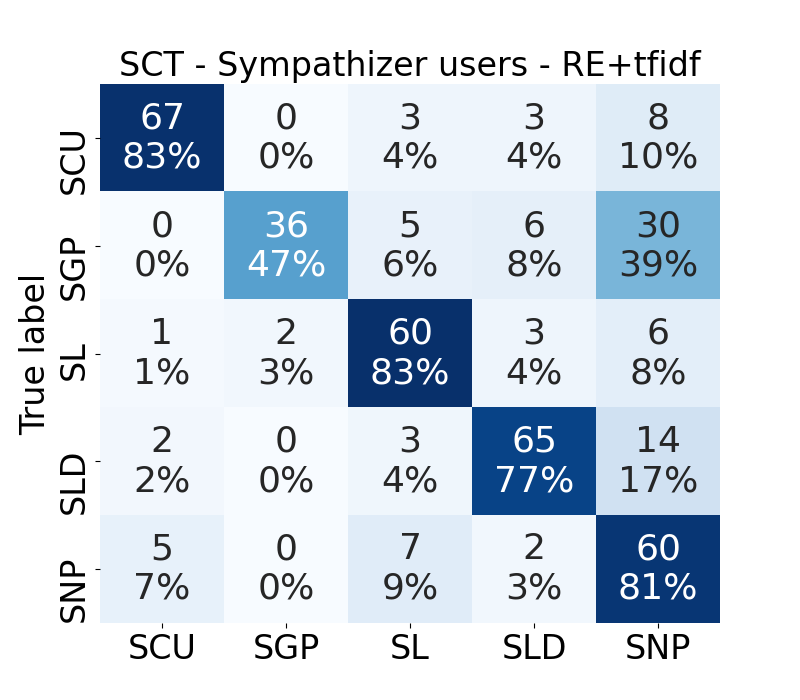}
    \caption{\footnotesize Confusion matrices of SCT Sympathizer users trained with RE (left) or RE+tfidf (right) representations.}
    \label{fig:hm_sct_sym} \hfil
\end{figure}

RE+tfidf representations achieve a macro-F1 score that is more than 10 points higher than RE representations on their own for SCT Sympathizer users. Further analysis of the confusion matrices for RE and RE+tfidf representations (Figure \ref{fig:hm_sct_sym}) reveals that users from other parties are often classified as SNP in RE representations, which is understandable considering the SNP’s prominence in the region. Specifically, SGP users are consistently misclassified as SNP users. However, when the RE+tfidf model is utilized, the classification becomes more accurate. The RE+tfidf model enhances the accuracy of the classification process, despite certain SGP Sympathizer users remain misclassified as SNP. That misclassification happens between users of two close parties, as both parties have ideological confluences and they govern together in the Scottish Parliament.

\begin{figure}[ht!]
\centering
    \includegraphics[width=0.38\linewidth]{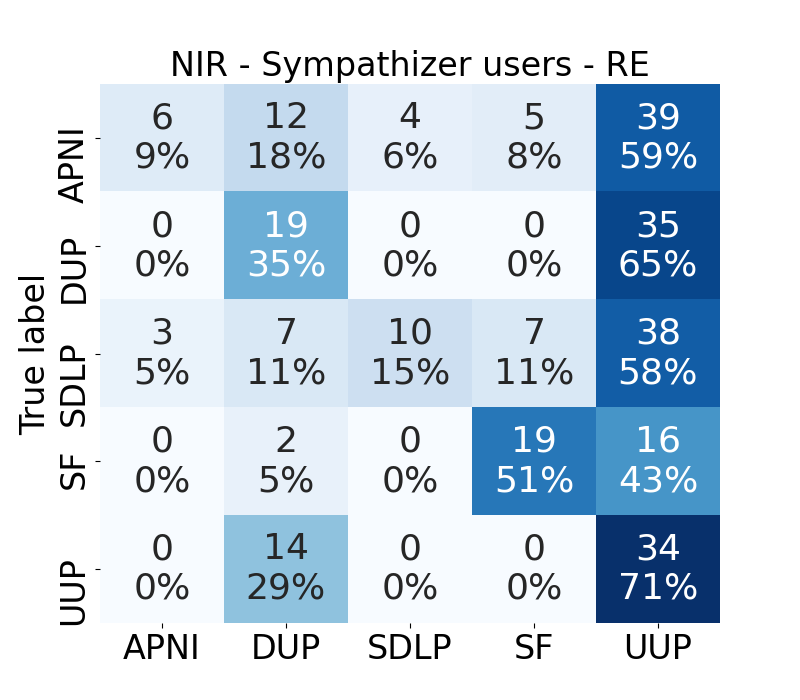}\hfil 
    \includegraphics[width=0.38\linewidth]{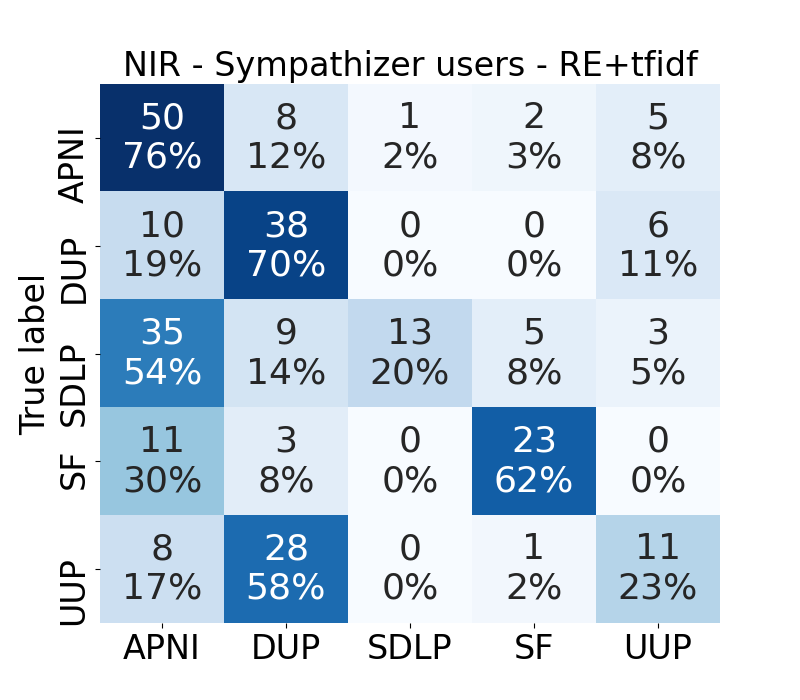}
    \caption{\footnotesize Confusion matrices of NIR Sympathizer users trained with RE (left) or RE+tfidf (right) representations.}
    \label{fig:hm_nir_sym} \hfil
\end{figure}

In terms of the performance of RE and RE+tfidf representations on NIR Sympathizer users, the inclusion of tfidf text representations results in a significant improvement of over 15 points in the F1 macro score. When examining the confusion matrices for both type of representations (Figure \ref{fig:hm_nir_sym}), several differences can be observed. The RE representations predominantly classify users as UUP users (Figure \ref{fig:hm_nir_sym} left), whereas the RE+tfidf representations primarily classify users as APNI users (Figure \ref{fig:hm_nir_sym} right). Moreover, the RE+tfidf representations struggle to identify UUP users as DUP users, which is an issue also present in the RE representations. One similarity between the two representations is that both perform better at classifying SF Sympathizer users compared to users from other parties, while facing difficulties in classifying SDLP Sympathizer users. Specifically, the RE+tfidf representations tend to misclassify SDLP users as APNI users and UUP users as DUP users. These failures can be attributed to the ideological proximity of these parties, making the RE+tfidf approach more accurate than the RE representations.

\paragraph{\textbf{Data visualization}} 
Combining RE interaction representations with tfidf textual representations achieves considerably better or similar results across all datasets especially enhancing the results for users that are less attached. In order to better understand and explain the effectiveness of the RE+tfidf HTIM representations, we visualize Member, Supporter and Sympathizer users datasets for the three regions, SCT, WAL and NIR by performing t-SNE dimensionality reduction into 2 dimensions.

\begin{figure}[ht!]
\centering
    \includegraphics[width=0.33\linewidth]{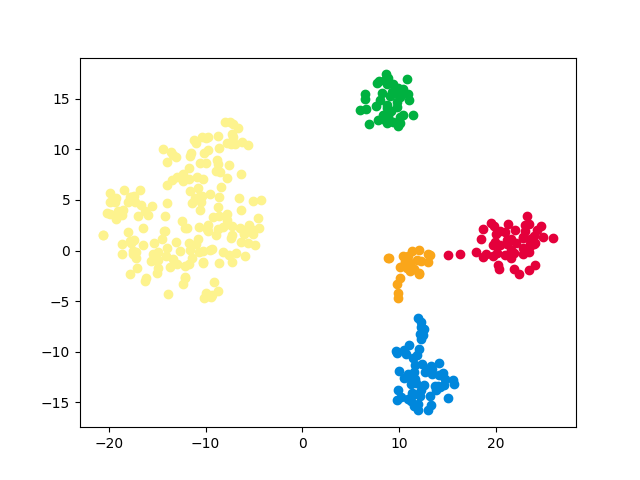}\hfil 
    \includegraphics[width=0.33\linewidth]{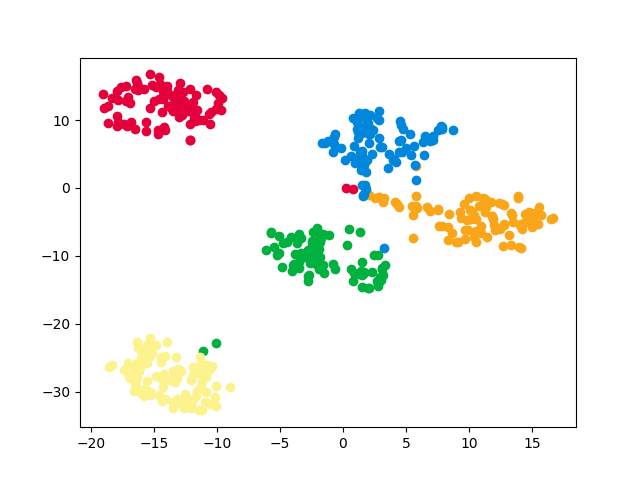}\hfil 
    \includegraphics[width=0.33\linewidth]{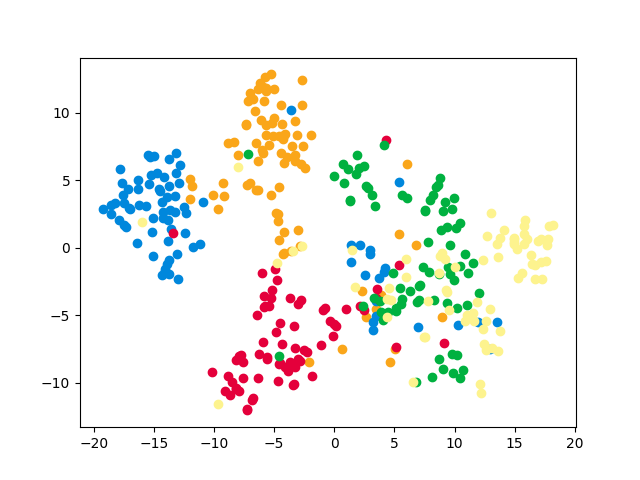}
    \caption{\footnotesize Visualization of t-SNE 2 dimension reduction of RE+tfidf HTIM representations for SCT Member (left), Supporter (center) and Sympathizer (right) user datasets.}
    \label{fig:emb_vis_sct} 
\end{figure}

\hfill \break
\noindent\textit{\textbf{SCT} (Figure \ref{fig:emb_vis_sct}):} In the Members dataset (Figure \ref{fig:emb_vis_sct} left), the different users are clearly grouped and defined by party. The SNP (\textcolor{snp}{$\bullet$}) and SGP (\textcolor{sgp}{$\bullet$}) are isolated and separated from the other parties, while the unionist parties SL (\textcolor{lab}{$\bullet$}), SLD (\textcolor{lib}{$\bullet$}), and SCU (\textcolor{con}{$\bullet$}) are clustered together. The positioning of the pro-independence parties (SNP and SGP) and the unionist parties (SL, SLD, and SCU) in different locations on the chart indicates a high polarization on the issue of national identity. Within the unionist community, the SLD serves as a link between SCU and SL, occupying a central political position. The Supporter dataset chart (Figure \ref{fig:emb_vis_sct} center) displays less attached users, but still tight-knit communities can be observed. The representation places the SGP between the SNP and SLD, indicating a more central position of SGP party among supporter users. The representation of Sympathizer users (Figure \ref{fig:emb_vis_sct} right) becomes sparser, with less homophily around political parties. The SCU occupies one extreme of the plot, while the SLD and SL take central positions. On the other extreme, the SNP and SGP are mixed, with the SGP closer to central positions. Despite the government treaty and the close alignment between SNP and SGP, the mixture between both parties is evident among Sympathizer users, demonstrating a significant level of closeness as observed in the error analysis.

\begin{figure}[ht!]
\centering
    \includegraphics[width=0.33\linewidth]{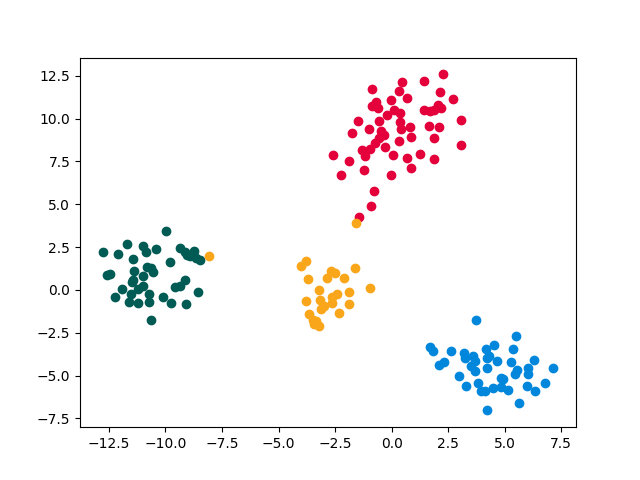}\hfil 
    \includegraphics[width=0.33\linewidth]{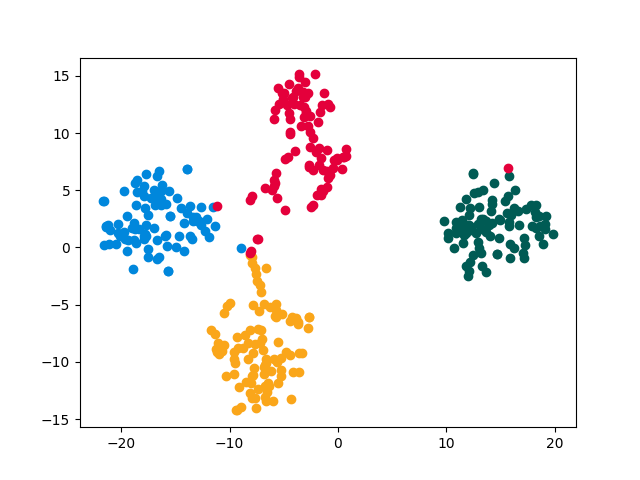}\hfil 
    \includegraphics[width=0.33\linewidth]{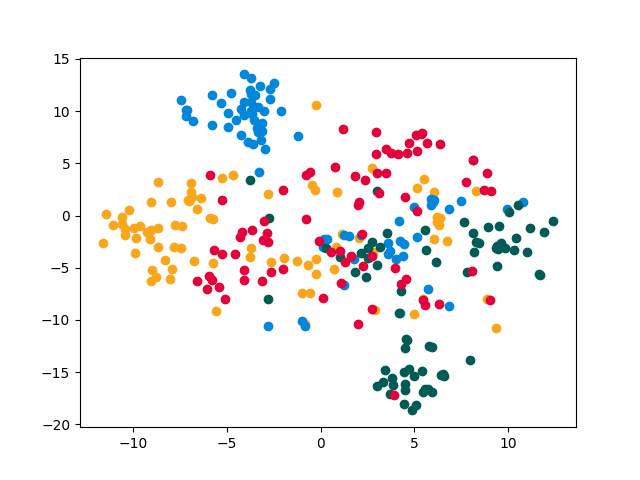}
    \caption{\footnotesize Visualization of t-SNE 2 dimension reduction of RE+tfidf HTIM representations for WAL Member (left), Supporter (center) and Sympathizer (right) user datasets.}
    \label{fig:emb_vis_wal} 
\end{figure}

\hfill \break
\noindent\textit{\textbf{WAL} (Figure \ref{fig:emb_vis_wal}):} Regarding Member users plot (Figure \ref{fig:emb_vis_wal}, left), the WLD (\textcolor{lib}{$\bullet$}) party occupies a central position, as it is considered ideologically situated at the political center. The remaining parties are positioned in the periphery of the diagram, with WL (\textcolor{lab}{$\bullet$}) located between PC (\textcolor{pc}{$\bullet$}) and WC (\textcolor{con}{$\bullet$}). The visualization of Supporter users (Figure \ref{fig:emb_vis_wal} center) shows that pro-independence PC users are isolated while the unionist WLD, WL and WC parties form their own cluster with labour (WL) and liberal (WLD) parties taking central positions. This configuration is similar to the SCT representation, with two prominent stances as pro-independence versus unionist. The representation of Sympathizer users (Figure \ref{fig:emb_vis_wal}, right) becomes sparser, similar to SCT, losing the clear distinction between the various communities. PC and a combined group of WLD and WC occupy opposite extremes of the plot, while WL appears to remain in the center, highlighting the political similarities from the perspective of sympathizers.

\begin{figure}[ht!]
\centering
    \includegraphics[width=0.33\linewidth]{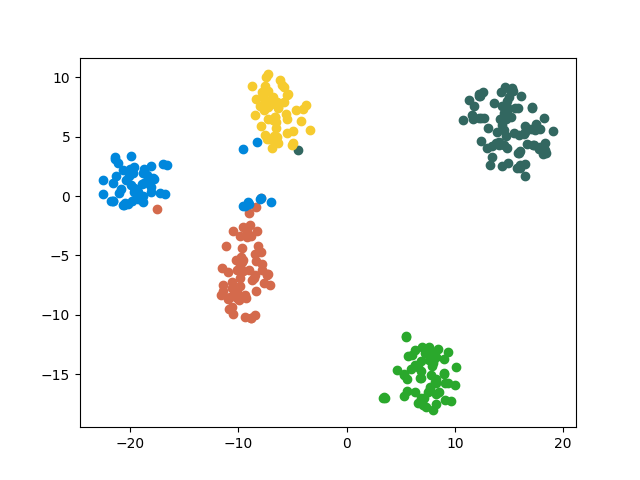}\hfil 
    \includegraphics[width=0.33\linewidth]{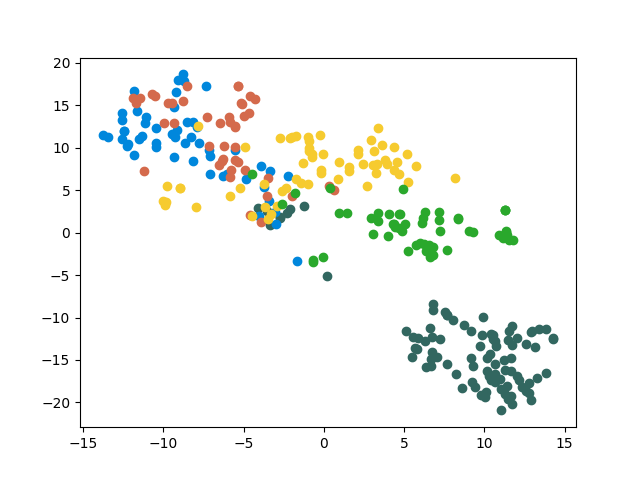}\hfil 
    \includegraphics[width=0.33\linewidth]{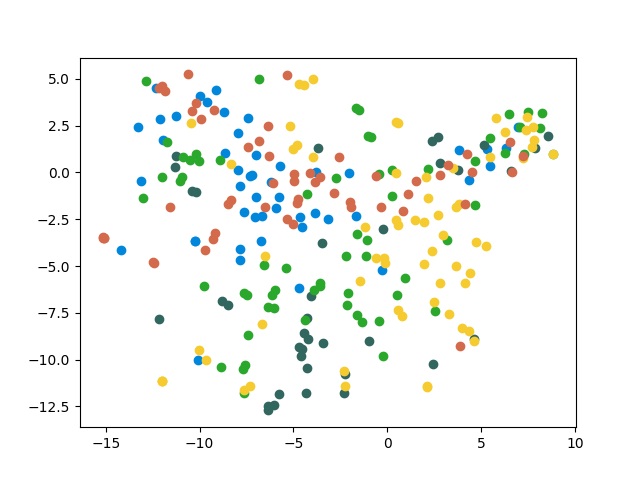}
    \caption{\footnotesize Visualization of t-SNE 2 dimension reduction of RE+tfidf HTIM representations for NIR Member (left), Supporter (center) and Sympathizer (right) user datasets.}
    \label{fig:emb_vis_nir} 
\end{figure}

\hfill \break
\noindent\textit{\textbf{NIR} (Figure \ref{fig:emb_vis_nir}):} A closer examination of the political party affiliations of Member users (Figure \ref{fig:emb_vis_nir} left) reveals that DUP (\textcolor{dup}{$\bullet$}), UUP (\textcolor{con}{$\bullet$}) and APNI (\textcolor{apni}{$\bullet$}) are positioned each in their own cluster characterized by liberal-conservative, right-wing, and unionist ideologies. On the other side of the graph, SF (\textcolor{sf}{$\bullet$}) and SDLP (\textcolor{sdlp}{$\bullet$}) occupy a distinct cluster with left-wing and pro-Irish orientations. When considering the representation of Supporter users (Figure \ref{fig:emb_vis_nir} center), significant polarization is observed between SF and the combined group of DUP and UUP, while APNI and SDLP assume more central positions. The visual representation effectively captures the ideological disparities, including the divisions between left and right orientations as well as pro-Irish and unionist perspectives. It is evident that the political parties aligned with unionist and right-wing ideologies are far apart from those with pro-Irish and left-wing leanings. In the representation of Sympathizer users (Figure \ref{fig:emb_vis_nir} right), the data becomes too sparse to extract any meaningful information. APNI sympathizers appear scattered throughout the plot, while UUP and DUP sympathizers occupy a similar space, confirming the challenges highlighted in the error analysis.

\hfill \break
While user representations for the Members dataset are clearly grouped, political party communities lose homophily when political attachment decreases and become sparser among Supporter and Sympathizer users datasets. It is noticeable that users not only are grouped depending on their assigned political party, but also that parties are positioned depending on their political similarities specially when communities are less defined. This phenomenon occurs for every annotated political party and for the three selected regions, meaning that the analyzed representations are able to consistently capture political information across different frameworks.

\section{Conclusion and Future Work}

In this work, we are the first to delve into the ability to predict the political leaning of social media users across different regions with multi-party systems using text and interactions as a data source. To achieve this, we propose Hybrid Text-Interaction Modeling (HTIM), a framework that enables integrating both data sources into the same model. To perform the experiments, we develop a new dataset spanning three UK regions, where we label users with different levels of political engagement (Members, Supporters and Sympathizers) with respect to the political party they align with. 

A look at the performance of each data source individually (i.e. text-based and interaction-based) shows that interactions are more effective in inferring political leaning than text-based approaches. However, these representations tend to underperform when dealing with users who have weaker political engagements.  To overcome this limitation, the use of our proposed HTIM to combine RE interaction representations with all the proposed textual representations results in considerable improvements across all datasets, especially with politically less attached users. The results are consistent across the regions and the different levels of political engagement, demonstrating its robustness. All in all, we demonstrate that our proposed HTIM achieves consistently improved performances, with a slight improvement on highly engaged users, but a remarkable improvement with those less engaged.

Considering the improvement obtained from the combination of textual and interaction data, we need to conduct further research to extract hybrid representations. We can experiment with various models using different datasets with missing information to address more realistic scenarios. As the collected datasets include interaction and textual data, we are able to try different configurations. Furthermore, we want to implement the proposed data extraction and user representation techniques in various other tasks, including hate-speech, disinformation or propaganda detection.

%\section*{Acknowledgments}

\end{document}